\documentclass[aps,prl,showpacs,showkeys,twocolumn]{revtex4-1} 
\usepackage{graphicx,amsmath,hyperref}
\usepackage{epstopdf}

\def\opone{\leavevmode\hbox{\small1\kern-3.8pt\normalsize1}}

\newcommand{\beq}{\begin{equation}}
\newcommand{\eeq}{\end{equation}}
\newcommand{\ba}{\begin{eqnarray}}
\newcommand{\ea}{\end{eqnarray}}
\newcommand{\bea}{\begin{eqnarray}}
\newcommand{\eea}{\end{eqnarray}}
\newcommand{\bma}{\begin{subequations}}
\newcommand{\ema}{\end{subequations}}
\newcommand{\bwt}{\begin{widetext}}
\newcommand{\ewt}{\end{widetext}}
\def\abs#1{|\,#1\,|}

\begin{document}

\title{Low loss light field modes in nano planar waveguide: An analytical approach based on the Rytov-Leontovich boundary conditions}

\author{N.M.Arslanov$^{1}$,  Ali A.Kamli$^{2}$, and S.A.Moiseev$^{1,3}$}
\email{narkis@yandex.ru, samoi@yandex.ru}
\affiliation{$^{1}$Kazan Quantum Center, Kazan National Research Technical University, 10 K. Marx, Kazan, 420111, Russia}
\affiliation{$^{2}$Department of Physics, Jazan University , Jazan 28824 P O Box 144 Saudi Arabia}
\affiliation{$^{3}$Zavoisky Physical-Technical Institute of the Russian Academy of Sciences, Kazan, Russia}


\begin{abstract}
We develop an analytical approach based on the Rytov-Leontovich boundary conditions to study light field modes in the nano planar waveguides.
The analytical results thus obtained demonstrate new conditions for optical low loss surface plasmon-polariton modes in the planar waveguides with dielectric-metamaterial interfaces.
The obtained results are confirmed with existing numerical calcultions. 
More general conditions for using the proposed analytical approach to study light field in the nano waveguides are also considered.
\end{abstract}

\date{\today}

\maketitle

\textit{Introduction.} Modern nano electronic  systems have reached spatial sizes which are comperable with few nano meters. Propagation of the signal pulses in such systems are characterized by considerable losses that significantly reduce the transfer rate and data processing. 
Nano optical systems can speed up the processing rate, but its large spatial sizes cause fundamental restrictions in matching with nano electronic systems. 
Some of these problems can be resolved with transfer to nano SPP (surface plasmon-polariton) waves at the interface between  metal and dielectric media \cite{Maier:2007:PFA, Raether:1988:SP, Genet-Ebbesen:2007:LTH, Agranovich-Mills:1982:SP}.

SPP waves \cite{Ritchie:1957:PLThF} demonstrate unique properties: they are highly confined near the interface and significantly slowed down compared with the electromagnetic waves. 
However, strong spatial confinement of SPP waves is accomponied by considerable ohmic losses that lead to the experimental constraints for deterministic manipulations of the SPP fields \cite{Berini:2009:LRSPP,Tame:2013:QP} and whence hinders its applications in quantum information technologies \cite{Tame:2013:QP,Nielsen-Chuang:2010:QCQI}.   
Strong supression of decoherence effects in SPP evolution requires searching for the most appropriate structure of the nano optical waveguides and its physical parameters for the implementation of SPP fields with sufficiently strong spatial confinement while at the same time suffer less losses. 

Using new metamaterial media \cite{Veselago:1967:EN, Pendry:2000:PL} opens new unusual properties of SPPs. 
Due to the destructive interference of magnetic and electric effects , the surface plasmon-polaritons can propagate over long distances  \cite{Kamli-Moiseev-Sanders:2008:CLLSP} on the metamaterial-dielectric interface with highly confined fields in the nanoscale spatial domain. 
Recently, it was shown that SPPs excited at waveguide structure (2D-interface of two media) can be used for considerable enhancement of usually weak photon-photon interactions \cite{Moiseev-Kamli-Sanders:2010:LLNP} that seem promising for the implementation of two- photon qubit gates \cite{NewJourPhys:2013:ROBUSTTOLOSS}, \cite{Siomau:2012:ECNIM} and for enhacements of various nonlinear light wave interactions \cite{Palomba-Novotny:2008}, \cite{Tan-Huang:2014:SuperluminalSPSolitons}.

So, it is of great interest to study the properties of SPPs in the various nano-sized waveguide structures to explore new opportunities in strengthening and controlling highly localized low loss light fields. 
This line of research was the subject of numerical simulations in recent works \cite{Lavoie-Leung-Sanders:2012:LLSMMWG}, \cite{pra-2013} devoted to SSP modes in the nano planar and cylindrical waveguides.

In this paper we develop an analytical approach for studying these light fields in nano planar waveguides by using the Rytov-Leontovich boundary conditions. 
We show that this approach allows the complicated mathematical expressions that appear in such complex spatial structures as planar nano waveguides to be reduced into more manageable form that lends itself to approximate analytical description of SPP light field modes in such spatial structures with high accuracy in the broadband spectral range. 
Finally based on the obtained results, we explore the appropriate parameters of nano optical planar waveguides with dielectric and metamaterial layers for the implementation of low loss SPP modes suitable for deterministic control.

\begin{figure}
\centerline{\includegraphics[scale=1]{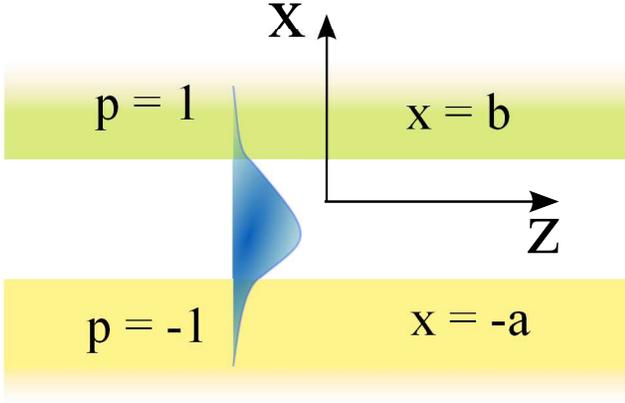}}
\vspace*{8pt}
  \caption{ (Color online) 
   The three-layer planar waveguide with walls made of a metamaterial and the dielectric inner layer, $a+b$ - the transverse dimension of the dielectric core. SPP 
modes propagate along the waveguide axis $z$.}
  \label{Figure1}
\end{figure}

\textit{The asymmetric planar waveguide (slab) basic equations}. The planar waveguide under consideration consists of three layers: a dielectric slab situated in the layer $(-a<x<b)$ and two asymmetric metamaterial claddings are in the two half spaces  $x<-a$ and $x>b$ as depicted in Fig. \ref{Figure1}.  
Both transverse electric (TE) and transverse magnetic (TM) SPP modes field can exist in such structure. The field components can be expressed through the Hertz potentials \cite{Born-Wolf:1999:PO}:

\begin{equation}
\begin{array}{l}
\psi_{p=\pm 1}= A_{p}e^{-pk_{p}x+i\theta},\\
\psi_{d}= \left(A_{d}\sin{g x}+B_{d}\cos{g x}\right) e^{i\theta},
\end{array}
\label{Equation-1_}
\end{equation}
\noindent
where $p (=\pm 1)$  is an index that describes the metamaterial medium (the subscript of the upper and lower layers), the index $d$ refers to the dielectric medium, $k_{p}$ is the transverse wavenumber in the layers of the metamaterial $k_{p}^2=h^2-k_{0}^{2}\varepsilon_{p} \mu_{p}$, $h$ is the wavenumber, parallel to the interface, $g$ is the transverse wavenumber in the dielectric, $h^2+g^2=k_{0}^{2}\varepsilon_{d} \mu_{d}$,  $\theta=hz-\omega t$, $\omega=2\pi c/\lambda$, $c$ is the speed of light, and $\lambda$ is its wavelength in the vacuum.

In this paper we restrict our discussion to TM SPP modes only (TE modes can be studied in a similar way) for which the  electric and magnetic components of the SPP modes are written as follows:

\begin{equation}
\begin{array}{l}
\vec{E}_{p}=\left( \frac{\partial^2 \psi_{p}}{\partial z \partial x},0,\frac{\partial^2 \psi_{p}}{\partial z^2}\right), 
 \vec{H}_{p}=\left(0, ik_{0}\varepsilon_{p} \frac{\partial \psi_{p}}{\partial x},0\right)
\end{array}.
\label{Equation-2_}
\end{equation}

From the boundary conditions for the fields at the two interfaces $x=-a$ and $x=b$:

\begin{equation}
\begin{array}{l}
E_{p}^{(z)}(-a;b)=E_{d}^{(z)}(-a;b),  \\
H_{p}^{(y)}(-a;b)=H_{d}^{(y)}(-a;b),
\end{array}
\label{Equation-2gr_}
\end{equation}

\noindent
we get the dispersion relation in the form of transcendental equation

\begin{equation}
\tan\left( g(a+b)-n\pi\right)=\frac{-\left( \frac{\varepsilon_{1}}{k_{1}}+\frac{\varepsilon_{-1}}{k_{-1}}\right) \frac{\varepsilon_{d}}{g}}{\left( \frac{\varepsilon_{d}}{g}\right)^2 -\frac{\varepsilon_{1}}{k_{1}} \frac{\varepsilon_{-1}}{k_{-1}}}
\label{Equation-3_}.
\end{equation}
 
\noindent
Solution of this transcendental equation allows determination of the phase and group velocities, transverse shape and propagation lenght of the TM SPP modes. However, Eq.(4) can not be solved analytically and one must resort to numerical methods. 
Numerical solutions have been previously obtained for certain parametres\cite{Adams:1981:IOW,  Lavoie-Leung-Sanders:2012:LLSMMWG, pra-2013}.

For metamaterial cladding one can use the typical behavior of the frequency dependent electric permittivity and magnetic permeability \cite{Lavoie-Leung-Sanders:2012:LLSMMWG}, \cite{pra-2013}: 
 
\begin{equation}
\begin{array}{l}
\varepsilon(\omega)/\varepsilon_{d}(\omega)=1-\omega_{e}^{2}/(\omega(\omega+i\gamma_{e}))\\
\mu(\omega)/\mu_{d}(\omega)=1+F\omega^2/(\omega_{0}^{2}-\omega(\omega+i\gamma_{m}))
\end{array},
\label{Equation-4_}
\end{equation}
\noindent
where the electric $\gamma_{e}$ and magnetic $\gamma_{m}$ damping rates are much less than the carrier frequency of interest; $\gamma_{e,m}\ll\omega$, where $\omega_{e}$ is a plasma frequency of the material, $\omega_{0}$ is a binding frequency, and F is a geometrical factor accounts for the magnetic oscillation strength. 
Spectral properties  of the refractive index $n_{p}(\omega)$ ($n^2_{p}(\omega)=\varepsilon (\omega)\mu(\omega)$) and impedance $ \zeta_{p}(\omega)$ ($ \zeta^2_{p}(\omega)=\mu(\omega)/ \varepsilon (\omega)$) are presented in Fig. \ref{Figure2}:

\begin{figure}[htdp]
\centerline{\includegraphics[scale=0.25]{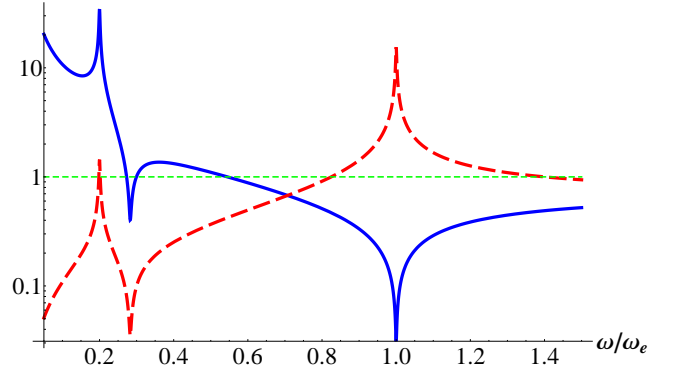}}
\caption{(Color online) Spectral behavior of the refractive index  $|n_{p}|$ (blue line), and $|\zeta_{p}|$  (red dashed line) of the metamaterial as the functions of  $\omega/\omega_{e}$.}
\label{Figure2}
\end{figure}

There are a series of intersting spectral domains for the metamaterial medium of interest characterized by its permittivity $\epsilon$ and permeability $\mu$ \cite{Veselago:1967:EN},\cite{Pendry:2000:PL} (Fig. \ref{Figure2}):

1) $\varepsilon(\omega)\approx|\varepsilon|$, $\mu(\omega)\approx|\mu|$, the refraction index Re$(n(\omega))>0$, for $(\omega>\omega_{e})$; here the metamaterial medium demonstrates the properties of a conventional dielectric medium.

2) $\varepsilon(\omega)\approx-|\varepsilon|$, $\mu(\omega)\approx-|\mu|$, the refraction index Re$(n(\omega))<0$. Here the light has a negative refraction angle in the metamaterial.

3)$\varepsilon\approx-|\varepsilon|$, $\mu\approx|\mu|$, $k\sim ik_{0}\sqrt{|\varepsilon| |\mu|}$ - $\varepsilon _{NM}$- \textit{«permittivity negative material»}. This indicates that only evanescent light modes can propagate in the metamaterial similar to light propagation in the usual metal.

4)$\varepsilon(\omega)\approx|\varepsilon|$, $\mu(\omega)\approx-|\mu|$,
$\mu_{NM}$- \textit{«permeability negative material»}. In this case, the light field  will also be evanescent modes in the metamaterial.

The cases 3) and 4) demonstrate the intermediate  properties in comparison to usual dielectric or left-handed materials where $\mu(\omega)\approx0$ or $\varepsilon (\omega)\approx0$.

As it is seen in Fig. \ref{Figure2}, there is a frequency area $\omega<0.25 \omega_{e}$ where  $\varepsilon \mu\gg 1$ is satisfied that makes applicable the Rytov-Leontovich boundary conditions  
\begin{equation}
\vec{E}_{(d)}\cong\zeta_{p}\left[\vec{n} \text{x} \vec{H}_{(d)}\right],
\label{Equation-RL_}
\end{equation}
 where $\vec{n}$ is the unit vector normal to the interface, instead of direct use of boundary conditions given in \eqref{Equation-2gr_}  (see, for example \cite{Yuferev:2008:SIBc}). 

By using the  Rytov-Leontovich boundary conditions \eqref{Equation-RL_}, we develop the pertubation theory with small parameter $|\zeta_{p}| <1$ for the analytical solution of the wave number in the transcendental equation \eqref{Equation-3_}. The  condition  $|\zeta_{p}| <1$  is satisfied for the frequency area $\omega<0.85 \omega_{e}$ (except small area around $\omega\cong  0.2 \omega_{e}$) (Fig. \ref{Figure2}) used below for the analytical calculations.
\begin{figure}
\centerline{\includegraphics[scale=0.25]{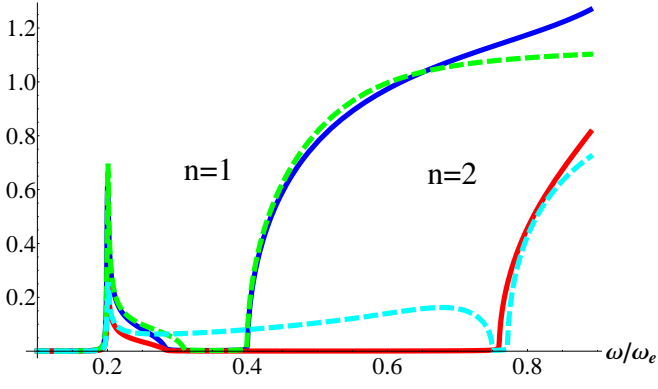}}
  \caption{ 
  The behavior of  $Re\{h\}$ for TM SPP modes:  $TM_{n=1}$, $TM_{n=2}$, 
  depending on the frequency for the analytical (solid curves)  and the numerical (dashed curves) solutions.  $2a=4\pi c/\omega_{\varepsilon}\approx 275$ nm.}
  \label{Figure3}
\end{figure}
\begin{figure}
\centerline{\includegraphics[scale=0.25]{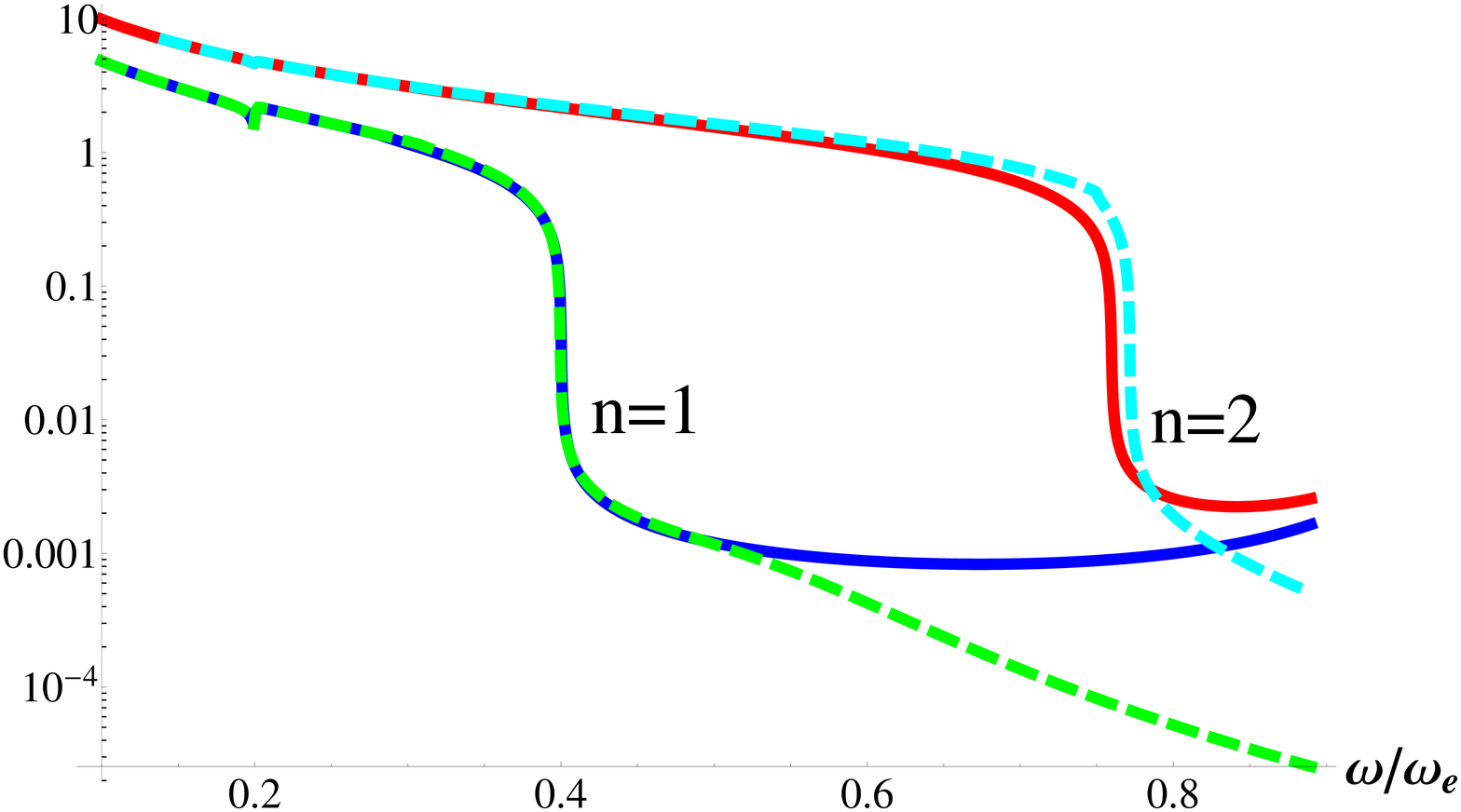}}
  \caption{ 
  The behavior of  $Im\{h\}$ for TM SPP modes:  $TM_{n=1}$, $TM_{n=2}$, 
  depending on the frequency for the analytical (solid curves)  and the numerical (dashed curves) solutions;  $2a=4\pi c/\omega_{\varepsilon}\approx 275$ nm.}
  \label{Figure4}
\end{figure}

\textit{The analytical approach.} The main idea of perturbation method is based on the series expansion  of the transverse wavenumber square in the slab region in powers of impedance \cite{Kacenelenbaum:1953}  ($\abs{\zeta_{p}}<1$):

\begin{equation}
\begin{array}{l}
g^2=g^2_{0}+\zeta g^2_{1}+...+\zeta^n g^2_{n}+...,\\ 
\psi=\psi_{(0)}+\zeta \psi_{(1)}+...+\zeta^n \psi_{(n)}+...\\
\end{array}
\label{Equation-5_}
\end{equation}

This approach was successfully applied for studying the light fields in the cylindrical tapering metal waveguides with dielectric core \cite{Kacenelenbaum:1953, Arslanov:2006:Probe, Arslanov-Moiseev:2007:PModes}.   
Here, we develop this approach by using the mode field presentation \eqref{Equation-1_}-\eqref{Equation-2_} and Rytov-Leontovich boundary equation \eqref{Equation-RL_} for the TM SPP modes, which lead to  the following analytical expression for the wave number in the asymmetric waveguide:

\begin{equation}
\begin{array}{llr}
 g^2={\left( \frac{\pi n}{b+a}\right)}^{2}-(\zeta_{1}+\zeta_{-1}) \frac{2ik_{0}\varepsilon_{d}}{b+a},\\
h^2=k_{0}^2 \varepsilon_{d}\mu_{d}-{\left( \frac{\pi n}{b+a}\right)}^{2}+(\zeta_{1}+\zeta_{-1})\frac{2ik_{0}\varepsilon_{d}}{b+a}.
\end{array}
\label{Equation-6_}
\end{equation}
\noindent
 For the symmetric waveguide ($\zeta_{1}=\zeta_{-1}=\zeta$ and  $b=a$) one can find the dispersion relation for even modes:

\begin{equation}
\begin{array}{llr}
 g^2={\left( \frac{\pi n}{a}\right)}^{2}-\zeta \frac{2ik_{0}\varepsilon_{d}}{a},\\
h^2=k_{0}^2 \varepsilon_{d}\mu_{d}-{\left( \frac{\pi n}{a}\right)}^{2}+\zeta \frac{2ik_{0}\varepsilon_{d}}{a},\\
\end{array}
\label{Equation-7_}
\end{equation}

\noindent
and for odd modes:
\begin{equation}
\begin{array}{llr}
 g^2={\left(\frac{\pi}{2a}+ \frac{\pi n}{a}\right)}^{2}-\zeta \frac{2ik_{0}\varepsilon_{d}}{a},\\
h^2=k_{0}^2 \varepsilon_{d}\mu_{d}-{\left(\frac{\pi}{2a}+ \frac{\pi n}{a}\right)}^{2}+\zeta \frac{2ik_{0}\varepsilon_{d}}{a},\\
\end{array}
\label{Equation-8_}
\end{equation}
\noindent
where  $n = 0,1,2, ...$ defines a set of the waveguide modes.

In Figs. \ref{Figure3},  \ref{Figure4}, we compare the approximate analytical solutions for TM even modes as given by equation \eqref{Equation-7_} with its exact numerical solutions obtained recently in \cite{Lavoie-Leung-Sanders:2012:LLSMMWG}, \cite{pra-2013} for the waveguide with transverse size $2a=4\pi c/\omega_{\varepsilon}\approx 275$ nm.
It is seen that the difference between the analytical and numerical solutions depends essentially on the frequency that is determined by spectral properties of the impedance and by the slab thickness. 
Each TM SPP mode has its own spectral domain where the good match between the analytical and numerical solutions occurs.  
  In particular, for the first TM SPP mode (n=1) , we have found that the analytical solution $h_{an}$ for the longitudinal wave number coinsides with the numerical one $h_{num}$ with precision $\frac{\abs{h_{an}}-\abs{h_{num}}}{\abs{h_{num}}}<10^{-2}$ for the spectral range 
  $\omega<0.25\omega_{e}$ (except small spectral range $\approx 0.04 \omega_{e}$ around $\omega\cong 0.2 \omega_{e}$ ) where the Rytov-Leontovich equation holds.
In the spectral range  $0.25 \omega_e<\omega<0.85\omega_e$, the precision of the wave number remains still high  enough $\frac{\abs{h_{an}}-\abs{h_{num}}}{\abs{h_{num}}}<5\cdot 10^{-2}$ (i.e. only 5 times smaller, except small area around cut-off frequency $0.4\omega_{e}$).

  It is interesting to note the drastic decrease of the   attenuation coefficient $Im{\{h\}}$ around the frequency $0.4\omega_e$. 
Low loss TM SPP modes can be exicted within the spectral domain which lies slightly higher than the cut-off frequency of this mode  $\omega>0.4\omega_e$  that reproduces numerical results of \cite{Lavoie-Leung-Sanders:2012:LLSMMWG, pra-2013}.
Similar  behaviour of the wave number and attenuation coefficient also occurs for the second TM-mode (n=2) close to its cut-off frequency $\omega\approx 0.77 \omega_e$ as it is seen in Figs.  \ref{Figure3}, \ref{Figure4}.  

\begin{figure}
\centerline{\includegraphics[scale=0.25]{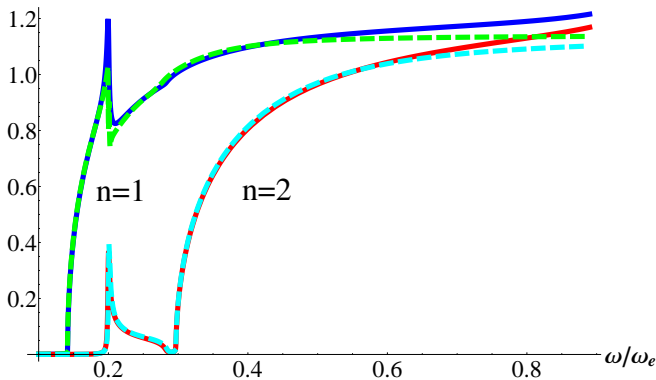}}
  \caption{ (Color online) The behavior of  $Re\{h\}$ for TM SPP modes:  $TM_{n=1}$, $TM_{n=2}$, 
  depending on the frequency for the analytical (solid curves)  and the numerical (dashed curves) solutions.  $2a=4\pi c/\omega_{\varepsilon}\approx 800$ nm.}
  \label{Figure5}
\end{figure}

\begin{figure}
\centerline{\includegraphics[scale=0.25]{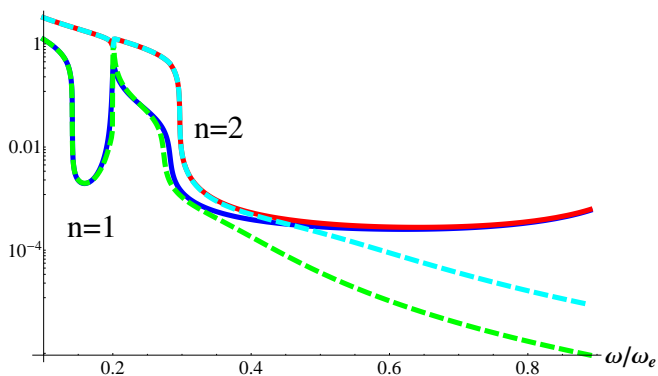}}
  \caption{ (Color online) The behavior of  $Im\{h\}$ for TM SPP modes:  $TM_{n=1}$, $TM_{n=2}$, 
  depending on the frequency for the analytical (solid curves)  and the numerical (dashed curves) solutions;  $2a=4\pi c/\omega_{\varepsilon}\approx 800$ nm.}
  \label{Figure6}
\end{figure}

Figs.\ref{Figure5}, \ref{Figure6} demonstrate the analytical and numerical calculations of the wave number and attenuation coefficent for larger thickness of the dielectic layer.  
The difference between these analytical and numerical results is negligible in the scale of these figures and precision increases for larger layer thickness. 
We have found from analysis of the solutions depicted in Fig. \ref{Figure6}, that larger thikness of the dielectric core leads to lower  losses for the light field modes within quite narrow spectral range near to $\omega=0.17 \omega_e$. 
The losses suppression is provided by the strong pushing of the field modes into the dielectric core from the metamaterial layers. Thus,  this effect observed earlier in \cite{Kamli-Moiseev-Sanders:2008:CLLSP} for 2D-interface can be also realized in the more complicated nano optical waveguides.        
It is evident that the discussed complicated spectral properties of the wave number and attenuation coefficient determined by the transcendental equation \eqref{Equation-3_} can be described  with high accuracy by the approximate analytical solutions \eqref{Equation-6_}-\eqref{Equation-8_} in large spectral range.


\textit{Conclusion}. In the present paper, we developed an analytical approach for description of the light field modes in the nano-optical asymmetric waveguide. We predicted new possibilities of low loss light field modes in the nano waveguides at some special parameters of the dielectic and metamaterial layers.  
The obtained analytical solutions are in good agreement with the recent numerical results \cite{Lavoie-Leung-Sanders:2012:LLSMMWG}, \cite{pra-2013} obtained for the planar waveguide  only with certain parameters of the metamaterial and dielectric layers.  
It should be noted that our analytical results hold with high accuracy within wider frequency range than it was assumed initially  on the basis of Rytov-Leontovich equation and of the refractive index relation.
This observation opens new promising possibilities for the analytical studies of light fields in various diverse nano optical structures with usual and artificial materials.      

N.M.A. and S.A.M. thank the Russian Scientiﬁc Fund through the grant no. 14-12-01333 for financial support of this work.

\bibliographystyle{apsrev4-1}
\bibliography{bib}

\end{document}